\documentclass[runningheads]{llncs}
\usepackage[T1]{fontenc}
\usepackage{graphicx}
\usepackage{amsmath}
\usepackage{amssymb}

\begin{document}
\title{SWR-Viz: AI-assisted Interactive Visual Analytics Framework for Ship Weather Routing }
\titlerunning{SWR-Viz}
%
\author{Subhashis Hazarika\inst{1} \and
Leonard Lupin-Jimenez\inst{1,2} \and
Rohit Vuppala\inst{1} \and
Ashesh Chattopadhyay\inst{2} \and
Hon Yung Wong\inst{1}}
\authorrunning{Hazarika et al.}
%
\institute{Fujitsu Research of America Inc., USA \and
University of California Santa Cruz, USA}

\maketitle              
\begin{abstract}
Efficient and sustainable maritime transport increasingly depends on reliable forecasting and adaptive routing, yet operational adoption remains difficult due to forecast latencies and the need for human judgment in rapid decision-making under changing ocean conditions. We introduce SWR-Viz, an AI-assisted visual analytics framework that combines a physics-informed Fourier Neural Operator wave forecast model with SIMROUTE-based routing and interactive emissions analytics. The framework generates near-term forecasts directly from current conditions, supports data assimilation with sparse observations, and enables rapid exploration of \textit{what-if} routing scenarios. We evaluate the forecast models and SWR-Viz framework along key shipping corridors in the Japan Coast and Gulf of Mexico, showing both improved forecast stability and realistic routing outcomes comparable to ground-truth reanalysis wave products. Expert feedback highlights the usability of SWR-Viz, its ability to isolate voyage segments with high emission reduction potential, and its value as a practical decision-support system. More broadly, this work illustrates how lightweight AI forecasting can be integrated with interactive visual analytics to support human-centered decision-making in complex geospatial and environmental domains.

\end{abstract}
\section{Introduction}
Ship weather routing (SWR) aims to optimize vessel trajectories by leveraging sea weather and ocean surface information to enhance safety, efficiency, and overall performance~\cite{zis2020ship}. For maritime operators, this can help reduce fuel consumption, shorten voyage time, and lower operational risks, leading directly to cost savings and improved reliability. By avoiding rough weather and capitalizing on favorable conditions, ships can also reduce emissions and contribute toward meeting carbon reduction goals, which is increasingly important as the maritime industry faces mounting regulatory and market pressure to decarbonize~\cite{chapman2007transport}.  

Despite these benefits, implementing SWR remains challenging. Ocean conditions are more volatile and complex than atmospheric conditions over land, making accurate and timely wave forecasts a major hurdle. This uncertainty directly impacts route planning for long voyages and the ability to re-plan routes under changing sea states. Effective routing depends not only on advanced forecast models and routing algorithms but also on skilled human judgment to balance safety, efficiency, and compliance. Pre-planned routes often require mid-voyage adjustments due to evolving conditions, and access to high-resolution, near-term forecasts is not always available when needed. Limited connectivity at sea further complicates access to updated forecast products.  

To address these operational gaps, we introduce SWR-Viz, an interactive visual analytics framework that integrates a physics-informed AI-based wave forecast emulator with ship routing simulations. The system is designed for human-in-the-loop operation, offering instant forecast generation, on-demand re-routing, and interactive visual route analytics that quantify engine power demand and emissions measures for associated harmful greenhouse gases. Users can generate several-hour forecasts from current conditions using a lightweight Fourier Neural Operator~\cite{fno20} (FNO)-based emulator, then immediately compute optimized routes with SIMROUTE~\cite{simroute22}, an open-source routing engine that incorporates vessel parameters and operational constraints. To ensure forecast stability and physical consistency of key ocean surface dynamics we incorporate both hard and soft physics constraints during the training process. FNOs are inherently resolution- and grid-agnostic; once trained, they generalize across spatial resolutions without retraining, enabling forecasts on arbitrary meshes or sampling patterns. This capability allows our framework to generate up-to-date wave-height predictions from sparse, heterogeneous observational data that are more readily available in real time for forecast initialization and assimilation. 


Running AI inference onboard is computationally faster and cheaper than waiting for high-resolution updated forecasts from external weather services, providing an advantage for rapid decision-making under limited bandwidth conditions. Therefore, SWR-Viz enables rapid \textit{what-if} analyses, which we term \textit{digital rehearsals} in our framework, where users can interactively define avoidance zones and compare alternative routes to evaluate safety and emissions trade-offs. We evaluated our AI models and SWR-Viz framework along popular shipping corridors in the Japan Coast and Gulf of Mexico, demonstrating its applicability across diverse oceanic regions. Digital rehearsals scenarios were run on expert-specified routes to assess system usability for real-world route analysis.



\section{Related Work and Background}
\textbf{AI-assisted Visualization Systems:}
Recent advances in visual analytics (VA) have demonstrated the potential of coupling domain-specific machine learning models with interactive visualization systems to support informed decision-making~\cite{ml4va,vis4ml_wang24}. Graph neural networks have been used to emulate and explore ocean simulations~\cite{shi2022gnn}, while MLP-based surrogates have enabled visual reasoning in cellular dynamics~\cite{NNVA2019}. Systems like HAiVA~\cite{haiva23} illustrate how physics-informed AI models can drive scenario manipulation and interactive hypothesis testing. As state-of-the-art climate AI models such as GenCast~\cite{gencast25}, PrithviWxC~\cite{prithviwxc24}, and ClimaX~\cite{climax23} enable faster, high-fidelity forecasts, platforms like Geospatial Foundation Model Studio~\cite{ibm_gfm_2023} exemplify a growing shift toward embedding these models within user-facing visual computing systems to support real-world, downstream decision workflows. Our work extends this paradigm by applying a Fourier Neural Operator to maritime forecasting and routing, allowing users to interact directly with forecasts and routing decisions in real time.

\textbf{Ship Weather Routing and Emissions-Aware Planning:}
SWR has long been recognized as a practical method to enhance voyage efficiency while reducing fuel consumption and greenhouse gas emissions. Recent studies estimate that operational measures such as routing and speed optimization can lead to 4-27\% emissions reduction, depending on environmental conditions and vessel characteristics~\cite{visir24,faber2023ceship}. Specific case studies, such as the use of SIMROUTE, report up to 30\% emissions reduction during severe storm events on longer shipping routes~\cite{simroute22}. Given that zero-carbon fuels are projected to cost significantly more than conventional fossil fuels~\cite{visir24}, energy-aware routing tools are crucial for economically sustainable decarbonization. Existing routing simulations~\cite{simroute22,visir24} often leverage pathfinding algorithms (e.g., A*~\cite{astar68}) and environmental forecasts to optimize ship trajectories. However, most current SWR tools lack real-time interactivity and integration with state-of-the-art fast AI-based forecasting. SWR-Viz fills this gap by providing an interactive visual analytics framework that integrates rapid forecasting, human-in-the-loop scenario exploration, and emissions-impact assessment within a unified operational interface.


\section{Requirement Analysis}
\label{RA_section}
Throughout the development of this framework, we engaged with several industry partners in the maritime technology sector, including a leading global shipping operators. Regular technology sharing meetings helped us identify key operational challenges in SWR and define the requirements for our interactive framework:

\begin{itemize}
	
	\item[\textbf{R1}]\label{R1} Ability to generate local wave forecasts from current conditions without relying on external weather agency updates, enabling rapid decisions in changing sea conditions.

	\item[\textbf{R2}]\label{R2} Tight integration of wave forecasting and routing execution to streamline operational planning and analysis. 
 
    \item[\textbf{R3}]\label{R3} Support for user-defined avoidance zones to incorporate updated regulatory constraints and explore what-if scenarios.   
 
    \item[\textbf{R4}]\label{R4} Visual tracking and filtering of route details, like fuel consumption, emission analytics and safety metrics in finer details for a more informed decision making.
    
\end{itemize}

\section{SWR-Viz Framework}
\begin{figure}[t!]
\centering
    \includegraphics[width = 1\linewidth]{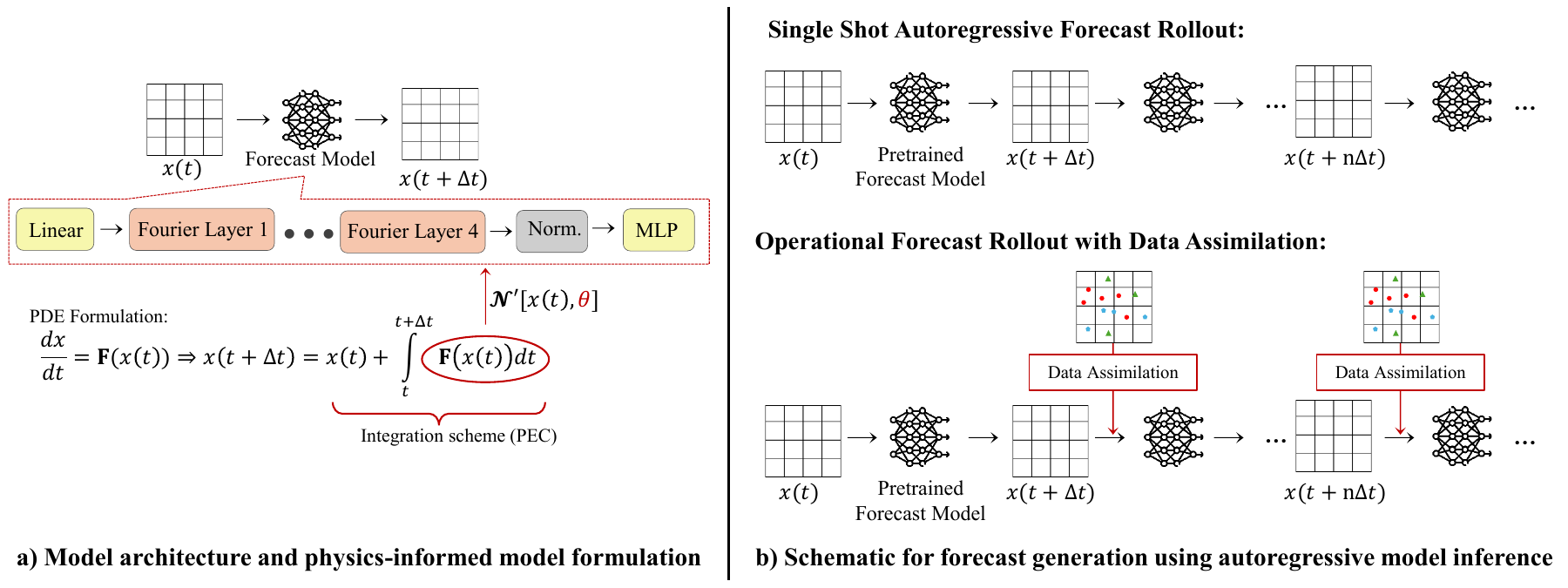} 
\caption{High-level illustration of our physics-informed AI wave forecast model}
\label{model_arch}
\end{figure}

Our proposed visual analytics framework comprises of i) a \textit{physics-informed AI wave forecast model}, ii) \textit{a ship routing simulation model} based on SIMROUTE, and iii) \textit{a front-end interactive visual analysis system} to interact with these models and facilitate digital rehearsals of ship routing scenarios.  


\subsection{AI-based Wave Forecast Model}
\label{aimodel_explain}
Our wave forecast model is based on Fourier Neural Operator (FNO) architecture~\cite{fno20}, a deep learning framework well-suited for modeling complex dynamical systems governed by partial differential equations (PDEs), such as those in weather and climate forecasting~\cite{oceannet24,fcds25,pathak2022fourcastnet}. Unlike conventional neural networks that operate on fixed-dimensional inputs, FNOs learn mappings between continuous spatio-temporal fields by projecting data into spectral space via the Fourier transform. This enables efficient global convolutions, resulting in faster training, fewer parameters, and better generalization for PDE-based systems.

As illustrated in Fig~\ref{model_arch}a, we model the evolution of sea surface wave fields $x(t)$ as a dynamical system governed by the first-order differential equation $\frac{dx}{dt} = F(x(t))$. The forecasting task is thus formulated as computing the future state $x(t+\Delta t)$ by integrating the learned dynamics over time: 
\begin{equation}
    x(t + \Delta t) = x(t) + \int_t^{t + \Delta t} F(x(t)) \, dt
\end{equation}
Here, \(F(x(t))\) is approximated by a FNO-based model, denoted \(N'[x(t), \theta]\), which captures the spatio-temporal wave dynamics from historical data. To numerically compute this integral, we adopt a Predictor-Error-Corrector (PEC) scheme as a hard physics constraint that provides a stable approximation of the solution:
\begin{equation}
    x(t+\Delta t) = x(t) + 0.5\Delta t \left( N'[x(t)] + N'[x(t) + N'[x(t)]] \right)
\end{equation}
This formulation helps embed physical structure into the learning process, allowing the model to produce stable and consistent forecasts over extended time horizons. To further guide the learning process with physical relevance, we introduce soft physics constraint as loss terms. An ocean-grid loss computes $L2$-errors exclusively over ocean regions, masking out land areas to focus learning where wave dynamics are meaningful. Additionally, a spectral loss minimizes reconstruction error in the 2D Fourier domain to preserve small-scale wave features and spatial coherence in the frequency space.

\textbf{Model Training:} We trained our models on three key wave variables necessary for SWR operations: \textit{Significant Wave Height} (VHM0), \textit{Mean Wave Direction} (VMDR), and \textit{Spectral Wave Peak Period} (VTPK). These were all obtained from the Copernicus Marine Service’s WAVERYS reanalysis product \cite{waverys_cmems}, a 3-hourly high-fidelity historical wave data repository. Since VMDR is a directional variable expressed in degrees, we transform it into two orthogonal components: VMDR$_x$ and VMDR$_y$, using its unit vector representation to avoid issues associated with learning in a cyclic angular space. This results in a total of four input channels for the wave state variable $x(t)$. The model comprised of 2.8 million parameters and took roughly 5 hours to train on a single NVIDIA A100 GPU machine for 27 years of historical wave data.

\textbf{Model Inference for Forecast Rollout:} Once trained, the AI wave forecast model can generate future states of the ocean surface in a physically consistent manner. As with conventional numerical forecasting systems, predictive accuracy gradually declines at longer lead times due to the accumulation of modeling errors. In operational settings, this limitation is typically mitigated through \textit{data assimilation}, where real-time sparse observations are used to periodically update and correct the forecast state. Leveraging the grid- and resolution-agnostic nature of FNOs, our model can seamlessly incorporate such sparse ocean observations to refine predictions without retraining, as illustrated in Fig.~\ref{model_arch}b. In our framework, this is achieved using a radial basis function (RBF) kernel density filter to adjust forecast fields toward observed values, ensuring stability and high fidelity over extended rollout periods. 


\subsection{Ship Routing Model}
To translate wave forecasts into actionable navigation strategies, we integrate our AI-based predictions with SIMROUTE~\cite{simroute22}, an open-source ship weather routing system that computes time-optimal routes using the A* pathfinding algorithm~\cite{astar68}. SIMROUTE incorporates vessel-specific inputs such as nominal cruising speed and applies varying wave effect on navigation models to account for speed reduction due to adverse sea states. These include empirical formulations based on wave height, encounter angle, ship length, and deadweight, providing flexibility to model a wide range of vessel types and conditions. The routing algorithm operates on a high-resolution navigation mesh, optimizing routes to minimize sailing time under dynamically evolving wave fields, and outputs both \textit{minimum-distance} and \textit{optimized routes} for comparison.

To generate emissions analytics, we processed the route details based on the STEAM2 methodology~\cite{steam2}, which estimates CO\textsubscript{2}, NO\textsubscript{x}, SO\textsubscript{x}, and particulate matter (PM) emissions by modeling the relationship between engine load, vessel speed, fuel consumption, and wave-induced power increase. Furthermore, the system includes safety metric evaluations following International Maritime Organization (IMO) guidelines~\cite{imo_safety}, detecting unstable navigation zones prone to phenomena like surf-riding and parametric rolling by comparing ship speed, wave characteristics, and vessel stability parameters. \textit{Surf-riding} occurs when a vessel accelerates uncontrollably on the steep front of following waves, risking broaching and loss of steering control. \textit{Parametric rolling} arises due to periodic variation in transverse stability, especially in head or following seas, and can lead to extreme roll amplitudes that threaten cargo and crew safety. 

Together with the AI-based wave forecast model, this routing module forms the computational backend of SWR-Viz, enabling seamless, integrated \textit{forecast–routing} pipelines that support interactive and exploratory visual analysis for operational SWR decision-making.
  
\subsection{Visual Analytics System}
\begin{figure*}[t!]
\centering
    \includegraphics[width = 0.95\linewidth]{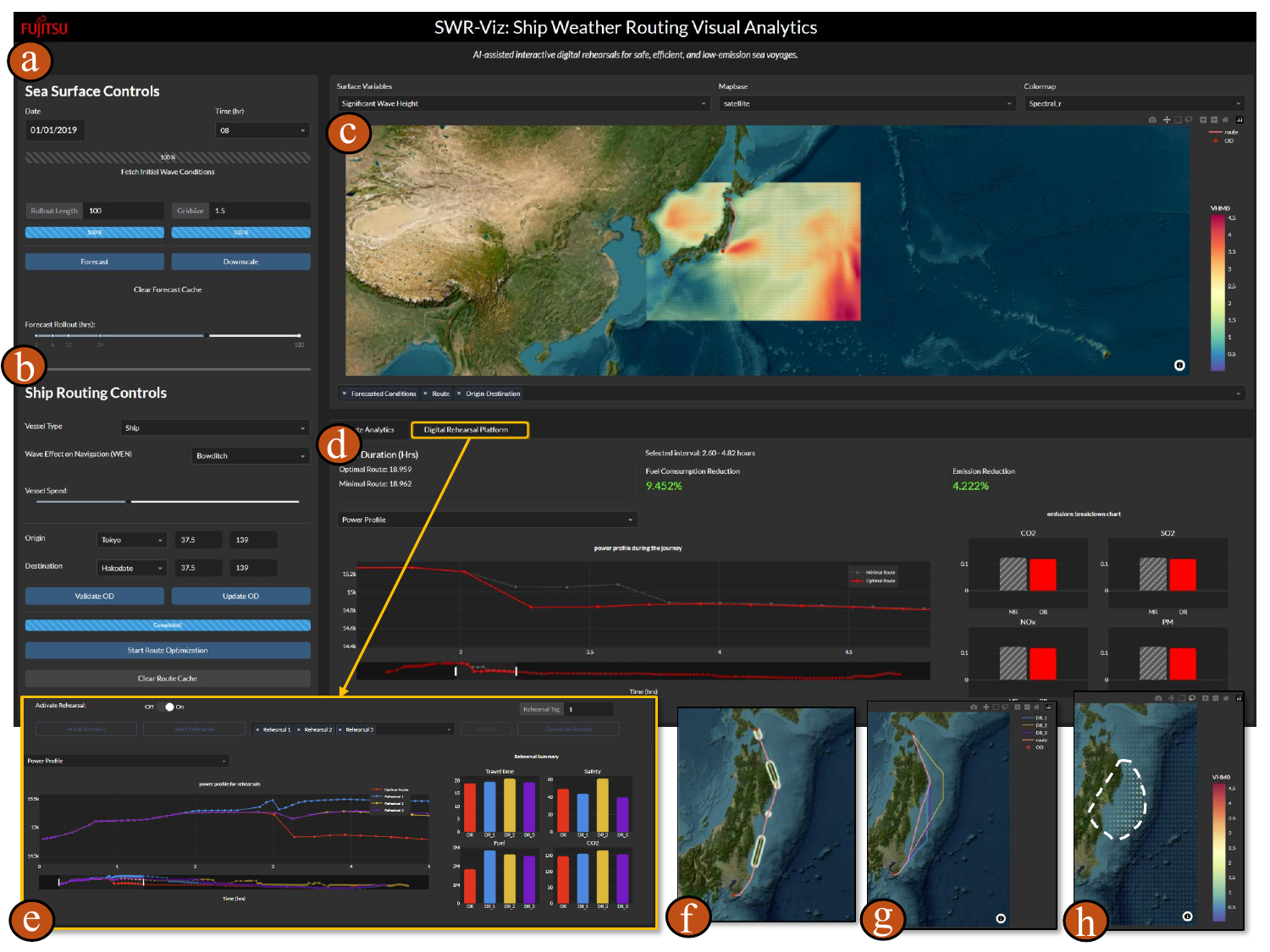} 
\caption{Overview of SWR-Viz, illustrating key VA components: (a) sea surface control panel, (b) ship routing control panel, (c) main visualization panel, (d) route analytics view, (e) digital rehearsal tab, (f) safety visualization, (g) comparative route rehearsal views, and (h) spatial constraint selection tools.}
\label{D-SWR}
\end{figure*}

The VA interface (Figure~\ref{D-SWR}) combines key interaction features and linked visualization panels to support the human-in-the-loop capabilities of SWR-Viz. Below, we outline the core VA capabilities in our system and how they address the desired requirements:

\textbf{Interaction Panels:}
The main interaction and control settings for the system are located on the left panel of the VA interface. Some specific controls related to spatial field visualization and digital rehearsal scenario exploration are grouped into individual analysis panels. Figure~\ref{D-SWR}(a) and (b) highlight the general settings for sea surface controls and ship routing controls, respectively.

In the sea surface control panel (Fig~\ref{D-SWR}a), users can select the date and time to define the initial condition for wave height information. This triggers the retrieval of relevant data from observational and/or reanalysis sources for the specified region of interest, which is then used as the initial condition for wave forecasting. The forecast button initiates an auto-regressive rollout of future wave conditions. This directly addresses the need for rapid local forecasting as outlined in \textbf{R1} in Section~\ref{RA_section}. Users can also define the forecast horizon and choose the grid resolution for the forecasted fields, which is particularly important for routing algorithms that require high-resolution inputs. The generated fields are rendered in the associated visualization panel (Figure~\ref{D-SWR}c), and a temporal rollout slider allows users to explore the evolving forecast data interactively.

The ship routing control panel (Fig~\ref{D-SWR}b) enables users to configure key parameters for the SIMROUTE module, including vessel type, average cruising speed, and the wave resistance model to apply. Origin and destination ports can be selected from a list of known ports or specified manually via latitude-longitude coordinates. For custom locations, the system provides validation tools to ensure the operability of the route based on the sea mask and forecast domain. Once location validation is completed, users can execute the routing algorithm to compute optimized paths and related analytics, fulfilling the integration requirement defined in \textbf{R2}.
 
\textbf{Visualization Panels:} Figure~\ref{D-SWR}c highlights the main visualization panel used to display geospatial data for both wave height fields and ship route planning details. This panel supports all the functional requirements outlined in Section~\ref{RA_section}. Users can visualize key wave parameters (VHM0, VMDR, VTPK) over selectable base map layers (by default, we use a satellite view for all figures presented in this paper). The interface allows users to choose the variable of interest, apply custom colormaps, and overlay multiple datasets using a multi-select dropdown menu. Once wave-optimized routes have been generated, their trajectories can be visualized by overlying on the forecasted wave fields. Safety indicators such as areas with high probability of parametric rolling or surf-riding can also be projected along the route path, providing spatial context to operational risk, as shown in Figure~\ref{D-SWR}f. When the \textit{digital rehearsal mode} is activated, this panel supports interactive spatial constraint definition (e.g., regulated or avoidance zones) as shown in Figure~\ref{D-SWR}h, enabling the comparison of alternative routing plans ( Figure~\ref{D-SWR}g) in what-if explorations.

Figure~\ref{D-SWR}d shows the analytics panel for evaluating optimized routes. It displays the engine power usage profile and the emissions breakdown for key pollutants, including CO\textsubscript{2}, NO\textsubscript{x}, SO\textsubscript{x}, and particulate matter (PM). Users can explore specific metrics using dropdown filters and zoom in on particular sections of the voyage via a temporal selection slider. This slider is linked to both the main map view (Figure~\ref{D-SWR}c) and the emission bar charts, allowing users to correlate selected segments of the journey with corresponding emission levels. Summary statistics, such as total travel time, fuel consumption, and percentage reduction in emissions achieved by the wave-optimized route compared to a baseline path, are also reported. Together, these features directly support the system’s goal of enabling informed, emissions-aware route planning, thereby fulfilling the requirements defined in \textbf{R4}.

\begin{figure}[t!]
\centering
    \includegraphics[width = 0.90\linewidth]{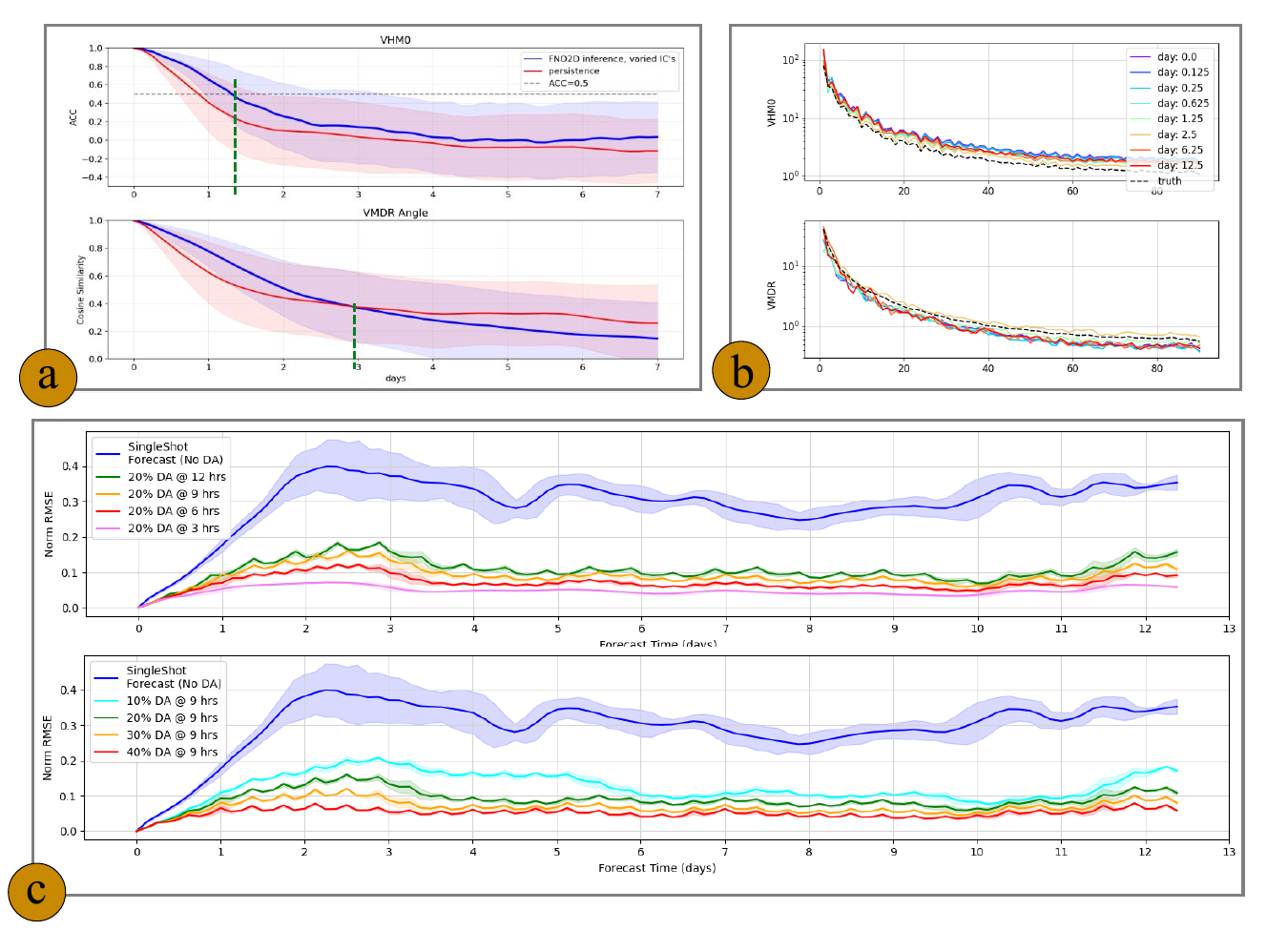} 
\caption{Validation studies for the AI-based wave forecast: (a) temporal stability measured by anomaly correlation for wave height and cosine similarity for wave direction against a persistence baseline, (b) numerical stability and physical consistency shown through spectral energy distribution across the first 100 wavenumbers compared to reanalysis data, and (c) operational forecast accuracy as normalized RMSE for single-shot rollouts and data assimilation strategies against ground-truth reanalysis.}
\label{validation_japan}
\end{figure}

\textbf{Digital Rehearsal Panel:} Another key requirement of the system is to support digital rehearsal of what-if scenarios, as specified by \textbf{R3}. To address this, we provide a dedicated tab within the VA system for digital rehearsal, shown in Figure~\ref{D-SWR}e, which includes specific interaction tools and comparative analytics for exploratory planning. Users can activate rehearsal mode and define spatial constraints using interactive tools such as a freeform lasso or rectangular selection on the geospatial visualization panel (Figure~\ref{D-SWR}h). This generates a high-resolution constraint field that modifies the routing domain for scenario testing. Once the constraint region is defined, users can re-run the routing algorithm to generate alternative paths that avoid the specified zones. These rehearsal routes can be visualized alongside the original optimal route in both the main map panel and the route analytics panel, enabling side-by-side comparison of key indicators (Figure~\ref{D-SWR}g). The system currently supports generation and analysis of up to five rehearsal paths, allowing users to compare different routing strategies under various operational constraints. For each scenario, metrics such as travel time, safety event percentage, fuel consumption, and emission reductions are presented in juxtaposed views. Finally, users have the option to save or export selected route scenarios for further evaluation and integration into broader decision-making workflows.

\section{Discussion}
\begin{table}[t!]
\centering
\caption{Comparison of routing performance across forecast settings.}
\label{tab:routing_results}
\resizebox{\textwidth}{!}{%
\begin{tabular}{|l|c|c|c|c|c|c|c|c|}
\hline
\textbf{Wave Height Data} & 
\textbf{Voyage Hours} & 
\textbf{Fuel (mT)} & 
\textbf{CO$_2$ (mT)} & 
\textbf{SO$_2$ (mT)} & 
\textbf{NO$_x$ (mT)} & 
\textbf{PM (mT)} & 
\textbf{Miles} & 
\textbf{Safety (\%)} \\
\hline
WAVERYS       & 19.69 & 57.30 & 150.14 & 0.46 & 3.03 & 0.08 & 469.69 & 55.00 \\
SingleShot Forecast  & 19.76 & 58.97 & 153.10 & 0.48 & 3.09 & 0.09 & 471.80 & 17.14 \\
20\% DA at 3hrs                & 19.69 & 57.31 & 150.20 & 0.46 & 3.02 & 0.08 & 469.70 & 55.00 \\
20\% DA at 6hrs                & 19.70 & 57.37 & 151.01 & 0.47 & 3.03 & 0.08 & 470.30 & 52.14 \\
20\% DA at 9hrs                & 19.69 & 57.37 & 151.22 & 0.47 & 3.03 & 0.08 & 470.62 & 49.40 \\
20\% DA at 12hrs               & 19.72 & 57.43 & 151.52 & 0.47 & 3.07 & 0.09 & 470.95 & 49.40 \\
\hline
\end{tabular}
}
\label{route_table}
\end{table}


\textbf{Validation Studies:} We evaluated our framework across two geographically distinct regions: the Japan coast (N $44^{\circ}$, S $25^{\circ}$, W $127^{\circ}$, E $163^{\circ}$) and the Gulf of Mexico (N $31^{\circ}$, S $17^{\circ}$, W $-99^{\circ}$, E $-74^{\circ}$). For each region, our AI-based wave forecast model was trained using 27 years of ground-truth reanalysis data from WAVERYS~\cite{waverys_cmems} for the period of 1993 to 2019 and validated on data from 2020 to 2022. We performed the following validation studies to evaluate the different elements of the framework from multiple aspects.
\begin{enumerate}
\item \textbf{Forecast Stability and Physical Consistency:}
To assess temporal stability, we compare our single-shot autoregressive forecasts against a persistence baseline, a common reference model that assumes current conditions remain unchanged into the future. While simple, this baseline provides a lower bound for skill and highlights how quickly predictive power degrades. We adopt two measures: the \textit{Anomaly Correlation Coefficient (ACC)} for significant wave height, which evaluates correlations between forecast and observed anomalies over time, and \textit{cosine similarity} for mean wave direction, which captures angular agreement between predicted and observed fields. These metrics were evaluated over 100 different initializations throughout 2020 to capture seasonal variability. As shown in Fig.~\ref{validation_japan}a, our forecasts retain ACC values above 0.5 for nearly 33 hours (green dashed line), substantially longer than persistence, while directional coherence remains strong for almost two days. Beyond anomaly-based skill scores, we also examine spectral stability to confirm physical consistency in the frequency domain (Fig.~\ref{validation_japan}b). Across all variables and lead times, the energy distribution over the first 100 wavenumbers closely matches that of the reanalysis ground truth, with no evidence of instability or high-frequency blow-up. This confirms both numerical stability, ensuring no spurious amplification in higher modes and resolution sensitivity, showing that spectral energy is preserved across scales. Overall, these results show that even in a single-shot autoregressive setting without data assimilation, the model produces forecasts that are both temporally stable and spectrally consistent over extended horizons. 

\item \textbf{Operational Forecast Accuracy:}  
In operational practice, single-shot forecasts are periodically updated through data assimilation when real-time observations become available (as explained in Section~\ref{aimodel_explain}). To evaluate the operational forecast quality of our model we performed data assimilation experiments at different time intervals and different rate of sparsity and compared against ground truth wave height information from WAVERYS reanalysis data. The top plot in Fig.~\ref{validation_japan}c shows normalized RMSE for a pure single-shot rollout (blue) against forecasts updated with 20\% sparse data at intervals of 3, 6, 9, and 12 hours respectively. We evaluated results from 100 different initial conditions across the year 2020 to capture the seasonal variance of sea condition across a year, which is reflected by the standard deviation bands in the plots. Assimilation consistently improves long-range accuracy, particularly at longer lead times.  We further tested robustness by fixing the assimilation interval at 9 hours while varying observation sparsity to replicate real-world conditions. The bottom plot in Fig.~\ref{validation_japan}c illustrates accuracy gains across 10–40\% data availability. These results highlight the grid- and resolution-agnostic design of the FNO-based forecast model, which can flexibly integrate sparse observations to produce high-quality forecasts across diverse operational scenarios.  

\item \textbf{Routing Performance:} To assess end-to-end route performance, we applied the framework to ship weather routing experiments using SWR-Viz. Following expert recommendations, we focused on the heavily used corridor between the Port of Tokyo and Hakodate. Table \ref{route_table} summarizes average route analytics over 10 runs from January 2022, with ship initialization parameters (average speed of 24 knots) kept fixed across all experiments. The results demonstrate that forecasts generated by our AI model, particularly the operational forecast with data assimilation closely approximate the routing outcomes derived from ground-truth WAVERYS reanalysis. Both emission-related measures (expressed in metric tons) and safety incident probabilities are well captured, providing experts with confidence that AI-based forecasts can support informed, emissions-aware routing decisions. 


\end{enumerate}

\textbf{Expert Feedback:} We engaged closely with a range of stakeholders during the development of SWR-Viz, facilitated by our business units. These included maritime experts from leading shipping companies in the Asia-Pacific region, sea navigation specialists, and senior corporate leaders. Though a formal structured user-study was not conducted within the scope of the project, the direct feedback we received was overwhelmingly positive, with experts particularly valuing the system’s usability and its ability to address practical challenges in ship weather routing. The human-in-the-loop visual analysis capabilities such as instant forecast generation, on-demand re-routing, and flexible specification of geographical constraints were highlighted as key differentiators, not currently available in conventional routing systems. They highlighted that, in general, energy savings and emission reductions for the complete voyage particularly for long routes are not significant when reported for the entire trip as a whole. However, by using the interactive visual selection features offered in SWR-Viz, it was possible to more conveniently isolate particular segments of the voyage with greater potential for energy efficiency, achieving reductions in the ranges of 8–10\%. They expressed interest in integrating selected features from SWR-Viz into existing platforms used in industry.

\section{Conclusion and Future Work}
In this paper, we presented SWR-Viz, an interactive visual analytics framework that integrates a physics-informed AI wave forecast model with a ship routing simulation engine. By enabling fast generation of wave forecasts, SWR-Viz supports rapid \textit{what-if} analyses and energy-efficient decision making for ship weather routing. As future work, we plan to extend the forecast model to include near-surface wind variables and to integrate key visual interaction features of SWR-Viz into existing industry platforms, as recommended by  experts.


\begin{credits}
\subsubsection{\ackname} This work was funded
by internal research funding of Fujitsu Research of America. A part of the work was also supported by NSF grant 2425667.

\subsubsection{\discintname}
The authors have no competing interests to declare that are
relevant to the content of this article.
\end{credits}
%
%
%
%






\bibliographystyle{splncs04}
\bibliography{template}

\end{document}